\documentclass[12pt]{article}
\usepackage{amsmath,epsfig,graphicx}
\input{epsf} 
\newcommand\as{\alpha_{\mathrm{S}}} 
\newcommand\f[2]{\frac{#1}{#2}} 
\def\beq{\begin{equation}} 
\def\eeq{\end{equation}} 
\def\beeq{\begin{eqnarray}} 
\def\eeeq{\end{eqnarray}} 
 
\def\to{\rightarrow}

\def\ptmin{p_{T{\rm min}}}
\def\ptmax{p_{T{\rm max}}}

\begin{document} 
\begin{titlepage}
\renewcommand{\thefootnote}{\fnsymbol{footnote}}
\vspace*{2cm}

\begin{center}
{\Large \bf Vector boson production at hadron colliders:}
\vskip 0.15cm
{\Large \bf a fully exclusive QCD calculation at NNLO }
\end{center}

\par \vspace{2mm}
\begin{center}
{\bf Stefano Catani${}^{(a)}$, Leandro Cieri${}^{(b)}$,
Giancarlo Ferrera${}^{(a)}$,\\
Daniel de Florian${}^{(b)}$}
and
{\bf Massimiliano Grazzini${}^{(a)}$}\\

\vspace{5mm}

${}^{(a)}$INFN, Sezione di Firenze and
Dipartimento di Fisica, Universit\`a di Firenze,\\
I-50019 Sesto Fiorentino, Florence, Italy\\

${}^{(b)}$Departamento de F\'\i sica, FCEYN, Universidad de Buenos Aires,\\
(1428) Pabell\'on 1 Ciudad Universitaria, Capital Federal, Argentina\\

\vspace{5mm}

\end{center}

\par \vspace{2mm}
\begin{center} {\large \bf Abstract} \end{center}
\begin{quote}
\pretolerance 10000

We consider 
QCD radiative corrections to the production of $W$ and $Z$ bosons
in hadron collisions. We present a fully exclusive calculation up to next-to-next-to-leading
order (NNLO) in QCD perturbation theory.
To perform this NNLO computation, we use a recently proposed version of the
subtraction formalism.
The calculation includes 
the $\gamma$--$Z$ interference, finite-width effects, the leptonic decay of the
vector bosons and the corresponding spin correlations.
Our calculation
is implemented
in a parton level Monte Carlo program. The program allows the user
to apply arbitrary kinematical cuts
on the final-state leptons and the associated jet activity, and 
to compute the corresponding
distributions in the form of bin histograms.
We show selected numerical results at the Tevatron and the LHC.

\end{quote}

\vspace*{\fill}
\begin{flushleft}
March 2009

\end{flushleft}
\end{titlepage}

\setcounter{footnote}{1}
\renewcommand{\thefootnote}{\fnsymbol{footnote}}

The production of $W$ and $Z$ bosons in hadron collisions 
through the Drell--Yan (DY) mechanism \cite{Drell:1970wh}
is extremely important for physics studies at hadron colliders.
These processes
have large production rates and offer clean experimental signatures, given the presence
of at least one high-$p_T$ lepton in the final state.
Studies of the production of $W$ bosons at the Tevatron lead to precise determinations
of the $W$ mass and width \cite{:2008ut}.
The DY process is also expected to provide standard candles for detector calibration
during the first stage of the LHC running. 

Because of
the above reasons, it is essential to have accurate
theoretical predictions for the vector-boson production cross sections
and the associated distributions.
Theoretical predictions with high precisions demand
detailed computations of radiative corrections.
The QCD corrections to the total cross section \cite{Hamberg:1990np}
and to the rapidity distribution \cite{Anastasiou:2003ds} of the vector boson 
are known up to the next-to-next-to-leading order (NNLO) in the strong coupling
$\as$. The fully exclusive NNLO calculation, including the leptonic decay 
of the vector boson, has been completed more recently \cite{DYdiff}.
Full electroweak corrections at ${\cal O}(\alpha)$
have been computed
for both $W$ \cite{ewW} and $Z$ production~\cite{ewZ}.

In this Letter we present a new computation of the NNLO QCD corrections
to vector boson production in hadron collisions.
The calculation includes
the $\gamma$--$Z$ interference, finite-width effects, the leptonic decay of the
vector bosons and the corresponding spin correlations.
Our calculation parallels the one recently completed
for Higgs boson production \cite{Catani:2007vq,Grazzini:2008tf},
and it is performed by using 
the same method.

The evaluation of higher-order QCD corrections to hard-scattering processes
is
complicated
by the presence
of infrared (IR) singularities at intermediate stages of the calculation that prevents
a straightforward implementation of numerical techniques.
Despite this difficulty, general methods have been developed in
the last two decades, which allow us
to handle and cancel IR singularities \cite{Giele:1991vf,Frixione:1995ms,Catani:1996vz} appearing in NLO QCD calculations.
In the last few years, several research groups have been working on extensions of
these methods to NNLO \cite{Kosower,Weinzierl,GGG,Frixione:2004is,ST}, and, recently, the NNLO calculation
for $e^+e^-\to 3~{\rm jets}$ was
completed by two groups \cite{threejets,Weinzierl:2008iv}.
Parallely, a new general method \cite{Anastasiou:2003gr},
based on sector decomposition \cite{sector}, has been proposed
and applied to
the NNLO calculations of $e^+e^-\to 2~{\rm jets}$ \cite{Anastasiou:2004qd}, 
Higgs \cite{Hdiff} and vector \cite{DYdiff} boson production in hadron collisions, and
to some decay processes \cite{Anastasiou:2005pn}.
Our method \cite{Catani:2007vq} applies to the production of colourless
high-mass systems in hadron collisions, and is based on
an extension of the subtraction formalism \cite{Frixione:1995ms,Catani:1996vz}
to NNLO that we briefly recall below.

We consider the inclusive hard-scattering reaction
\begin{equation}
h_1+h_2\to V(q)+X,
\end{equation}
where the collision of the two hadrons $h_1$ and $h_2$
produces the vector boson $V$ ($V=Z/\gamma^*, W^+$ or $W^-$), 
with four-momentum $q$ and high invariant mass $\sqrt {q^2}$. 
At next-to-leading order (NLO), two kinds of corrections contribute: i) {\em real} corrections,
where one parton recoils against $V$; ii) {\em one-loop virtual} corrections to
the leading order (LO) subprocess. Both contributions are separately 
IR divergent, but 
the divergences cancel in the sum.
At NNLO, three kinds of corrections must be considered: i) {\em double real} 
contributions, where two partons recoil against $V$; ii) {\em real-virtual} 
corrections, where one parton recoils against $V$ at one-loop order; 
iii) {\em two-loop virtual} corrections to the LO subprocess.
The three contributions are still
separately divergent, and the calculation has
to be organized so as to explicitly achieve the cancellation of the 
IR divergences.

We first note that, at LO, the transverse momentum 
$q_T$ of $V$ is exactly zero.
As a consequence, as long as $q_T\neq 0$, the (N)NLO contributions are actually given by the (N)LO 
contributions to $V+{\rm jet(s)}$.
Thus, we can write the 
cross section as
\begin{equation}
\label{Vplusjets}
d{\sigma}^{V}_{(N)NLO}|_{q_T\neq 0}=d{\sigma}^{V+{\rm jets}}_{(N)LO}
\;\; .
\end{equation}
This means that, when $q_T\neq 0$, the 
IR divergences in our NNLO calculation are those in 
$d{\sigma}^{V+{\rm jets}}_{NLO}$: they can be
treated by using
available NLO methods
to handle and cancel IR singularities 
(e.g., the general NLO methods in
Refs.~\cite{Giele:1991vf,Frixione:1995ms,Catani:1996vz}).
The only remaining singularities of NNLO type are associated to the limit 
$q_T\to 0$. Following Ref.~\cite{Catani:2007vq} we treat them by an additional subtraction. 
Our key point is that the singular behaviour 
of $d{\sigma}^{V+{\rm jets}}_{(N)LO}$ when $q_T\to 0$ is well known:
it comes out in the 
resummation program \cite{Catani:2000jh}
of logarithmically-enhanced contributions
to transverse-momentum distributions.
Therefore, the additional subtraction can be worked out by using a counterterm, $d{\sigma}^{CT}_{(N)LO}$,
whose 
general structure \cite{Catani:2007vq}
depends only on
the flavour of the initial-state partons involved in the LO
partonic subprocess ($q{\bar q}$ annihilation in the case
of vector-boson production, $gg$ fusion in the
case of Higgs boson production). 

Our extension of Eq.~(\ref{Vplusjets}) to include
the contribution at $q_T=0$ is \cite{Catani:2007vq}:
\begin{equation}
\label{main}
d{\sigma}^{V}_{(N)NLO}={\cal H}^{V}_{(N)NLO}\otimes d{\sigma}^{V}_{LO}
+\left[ d{\sigma}^{V+{\rm jets}}_{(N)LO}-
d{\sigma}^{CT}_{(N)LO}\right]\;\; .
\end{equation}
Comparing with the right-hand side of
Eq.~(\ref{Vplusjets}), we have subtracted
the (N)LO counterterm $d{\sigma}^{CT}_{(N)LO}$
and added a contribution at $q_T=0$, which is needed to 
obtain the correct total cross section.
The coefficient ${\cal H}^{V}_{(N)NLO}$ does not depend on $q_T$
and is obtained by the (N)NLO truncation of the hard-scattering perturbative function
\begin{equation}
{\cal H}^{V}=1+\f{\as}{\pi}\,
{\cal H}^{V(1)}+\left(\f{\as}{\pi}\right)^2
{\cal H}^{V(2)}+ \dots \;\;.
\end{equation}
According to Eq.~(\ref{main}), the NLO calculation  of $d{\sigma}^{V}$ 
requires the knowledge
of ${\cal H}^{V(1)}$ and the LO calculation of $d{\sigma}^{V+{\rm jets}}$.
The general (process-independent) form of 
the coefficient 
${\cal H}^{(1)}$ is known: the precise relation between 
${\cal H}^{(1)}$ and the IR finite part of the
one-loop correction to a generic LO subprocess is explicitly derived in 
Ref.~\cite{deFlorian:2000pr}.
At NNLO, the coefficient ${\cal H}^{V(2)}$ is also needed, together with the
NLO calculation of $d{\sigma}^{V+{\rm jets}}$.
The calculation of the general structure
of the coefficients ${\cal H}^{(2)}$ is in progress.
Meanwhile,
by using the available analytical results at ${\cal O}(\as^2)$ for the
total cross section \cite{Hamberg:1990np} and the transverse-momentum 
spectrum \cite{VqtNLO} of the vector boson,
we have 
explicitly computed
the coefficient ${\cal H}^{V(2)}$ of the DY process.
Since the NLO corrections $d{\sigma}^{V+{\rm jets}}_{NLO}$
to $q{\bar q}\to V+{\rm jet(s)}$ are also available 
\cite{Giele:1993dj}, using Eq.~(\ref{main})
we are able to complete our fully-exclusive NNLO calculation
of vector-boson production.

We have encoded our NNLO computation in a parton level
Monte Carlo program, in which
we can implement arbitrary IR safe cuts on the final-state
leptons and the associated jet activity.

In the following
we present an illustrative selection of
numerical results for $Z$ and $W$ production at the Tevatron and the LHC.
We consider $u,d,s,c,b$ quarks in the initial state.
In the case of
$W^\pm$ production,
we use the (unitarity constrained) CKM matrix elements $V_{ud}=0.97419$, $V_{us}=0.2257$, $V_{ub}=0.00359$,
$V_{cd}=0.2256$, $V_{cs}=0.97334$, $V_{cb}=0.0415$ from the PDG 2008 \cite{Amsler:2008zzb}. 
In the case of $Z$ production, 
additional Feynman diagrams with fermionic triangles 
should be taken into account.
Their contribution cancels out for
each isospin multiplet when massless quarks are considered. The effect of
a finite top-quark mass
in the third generation has been considered and found extremely small \cite{Dicus:1985wx},
so it 
is 
neglected in our calculation.
As for the electroweak couplings, we use the so called $G_\mu$ scheme,
where the input parameters are $G_F$ , $m_Z$, $m_W$. In particular we 
use the values
$G_F = 1.16637\times 10^{-5}$~GeV$^{-2}$,
$m_Z = 91.1876$~GeV, $\Gamma_Z=2.4952$~GeV, $m_W = 80.398$~GeV
and $\Gamma_W=2.141$~GeV.
We use the 
MSTW2008 \cite{Martin:2009iq} sets
of parton distributions, with
densities and $\as$ evaluated at each corresponding order
(i.e., we use $(n+1)$-loop $\as$ at N$^n$LO, with $n=0,1,2$). The 
renormalization and factorization scales are fixed to the value 
$\mu_R=\mu_F=m_V$, where $m_V$ is the mass of the vector boson.

\begin{figure}[htb]
\begin{center}
\begin{tabular}{c}
\epsfxsize=12truecm
\epsffile{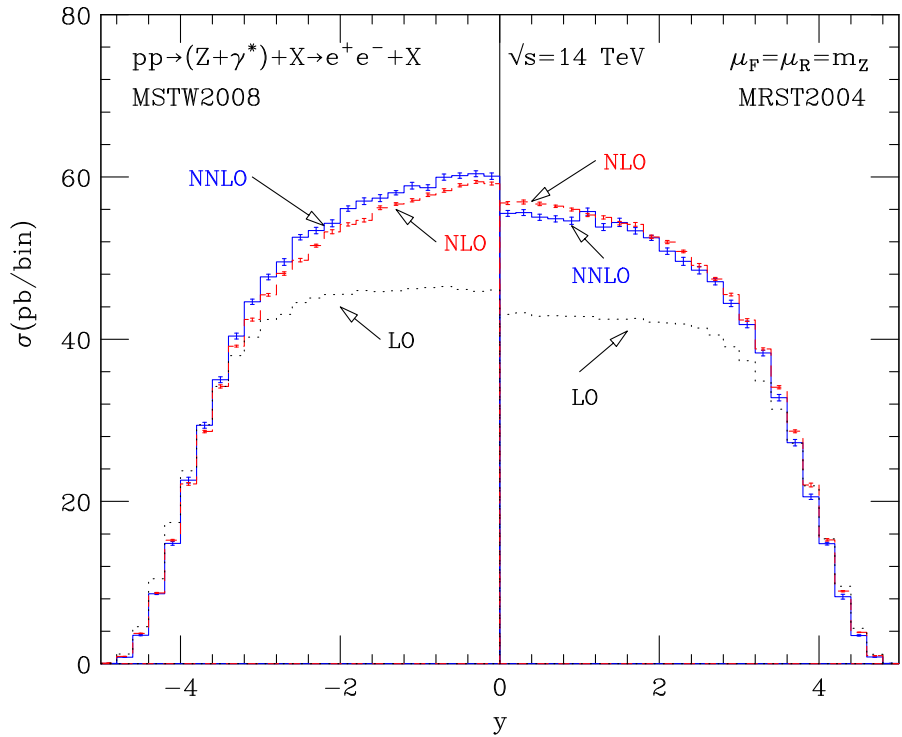}\\
\end{tabular}
\end{center}
\caption{\label{fig:y34lhc}
{\em Rapidity distribution of an on-shell $Z$ boson at the LHC. Results obtained
with the MSTW2008 set (left panel) are compared with those obtained with the MRST2004 set (right panel).}}
\end{figure}

We start the presentation of our results by considering
the inclusive production of $e^+e^-$ pairs from the decay of an on-shell $Z$ boson
at the LHC. In Fig.~\ref{fig:y34lhc} (left panel) we show the rapidity 
distribution of the $e^+e^-$ pair at LO, NLO and NNLO, computed 
by using
the MSTW2008 partons.
The corresponding cross sections\footnote{Throughout the paper, the errors
on the values of the cross sections and the error bars in the plots refer to an
estimate of the numerical errors in the Monte Carlo integration.}
 are $\sigma_{LO}=1.761 \pm 0.001$~nb, 
$\sigma_{NLO}=2.030 \pm 0.001$~nb and $\sigma_{NNLO}=2.089 \pm 0.003$~nb. 
The total cross section is increased by about $3$\%
in going from NLO to NNLO. In Fig.~\ref{fig:y34lhc} (right panel)
we also show 
the results obtained 
by using the 
MRST2002 LO \cite{Martin:2002aw} 
and MRST2004 \cite{Martin:2004ir}
sets of parton distribution functions. The corresponding cross sections are
$\sigma_{LO}=1.629 \pm 0.001$~nb, $\sigma_{NLO}=1.992 \pm 0.001$~nb and 
$\sigma_{NNLO}=1.954 \pm 0.003$~nb. In this case the total cross section 
is decreased by about $2$\%
in going from NLO to NNLO.

\begin{figure}[htb]
\begin{center}
\begin{tabular}{c}
\epsfxsize=12truecm
\epsffile{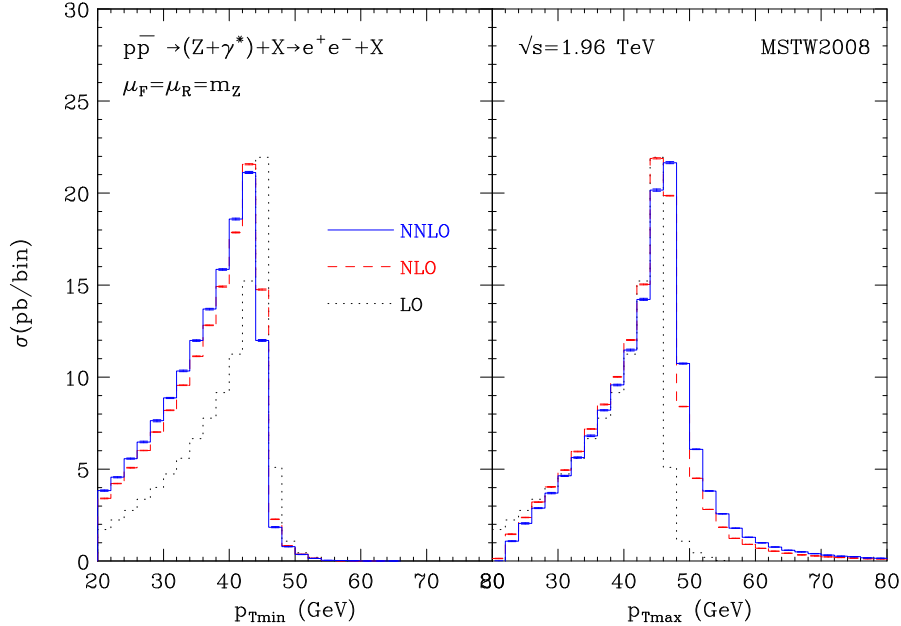}\\
\end{tabular}
\end{center}
\caption{\label{fig:ptminmax}
{\em Distributions in $\ptmin$ and $\ptmax$ for the $Z$ signal at 
the Tevatron.}}
\end{figure}

We next consider the production of $e^+e^-$ pairs from $Z/\gamma^*$ bosons at the Tevatron.
For each event, we classify the lepton transverse momenta according to their
minimum and maximum values,  
$\ptmin$ and $\ptmax$.
The 
leptons
are required to have a minimum $p_T$ of 20~GeV and pseudorapidity $|\eta|<2$.
Their invariant mass is required to be in the range 70~GeV~$<m_{\,e^+e^-}<110$~GeV.
The accepted cross sections are $\sigma_{LO}=103.37 \pm 0.04$~pb,
$\sigma_{NLO}=140.43 \pm 0.07$~pb and $\sigma_{NNLO}=143.86 \pm 0.12$~pb.
In Fig.~\ref{fig:ptminmax} we plot
the distributions in $\ptmin$ and $\ptmax$ at LO, NLO and NNLO.
We note that at LO the $\ptmin$ and $\ptmax$ distributions are kinematically
bounded by $p_T\le Q_{\rm max}/2$, where $Q_{\rm max}=110$~GeV is the maximum allowed invariant mass of the $e^+e^-$ pairs.
The NNLO corrections have a visible impact on the shape of the $\ptmin$ and $\ptmax$ distribution and
make the $\ptmin$ distribution softer, and the $\ptmax$ distribution harder.

We finally consider the production of a charged lepton plus missing $p_T$ through the decay of a $W$ boson ($W=W^+,W^-$) at the Tevatron.
The charged lepton is selected to have $p_T>20$~GeV and $|\eta|<2$ and the missing $p_T$
of the event should be larger than 25~GeV. We define the transverse mass of the
event as $m_T=\sqrt{2p_T^lp_T^{\rm miss}(1-\cos\phi)}$, where $\phi$ is the angle between the
the $p_T$ of the lepton and the missing $p_T$. The accepted cross sections are
$\sigma_{LO}=1.161 \pm 0.001$~nb,
$\sigma_{NLO}=1.550\pm 0.001$~nb and $\sigma_{NNLO}=1.586 \pm 0.002$~nb.

\begin{figure}[htb]
\begin{center}
\begin{tabular}{c}
\epsfxsize=12truecm
\epsffile{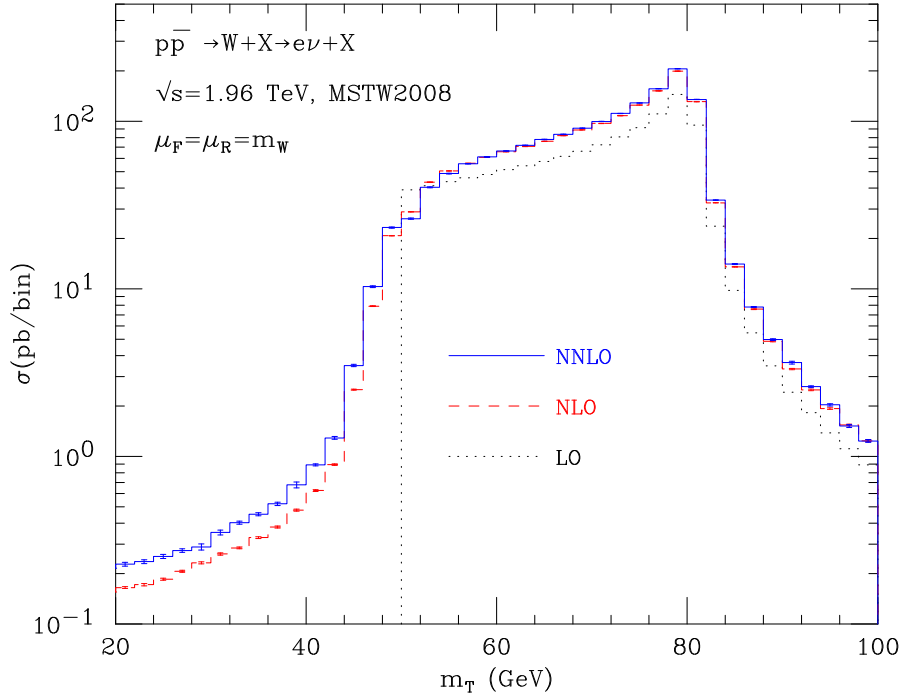}\\
\end{tabular}
\end{center}
\caption{\label{fig:tmass}
{\em Transverse mass distribution for $W$ production at the Tevatron.}}
\end{figure}

In Fig.~\ref{fig:tmass} we show the $m_T$ distribution at LO, NLO and NNLO.
We note that at LO the distribution has a kinematical boundary at $m_T=50$~GeV.
This is due to the fact that at LO the $W$ is produced with zero 
transverse momentum:
therefore,
the requirement $p_T^{\rm miss}>25$~GeV sets
$m_T\geq 50$~GeV. 
Around the region where $m_T=50$~GeV there are perturbative 
instabilities in going from LO to NLO and to NNLO.
The origin of these perturbative instabilities is 
well 
known 
\cite{Catani:1997xc}:
since the LO spectrum
is kinematically bounded by $m_T\geq 50$~GeV,
each higher-order perturbative contribution produces
(integrable) logarithmic singularities in the vicinity of
the boundary.
We also note that, below the boundary, the NNLO corrections to the NLO result
are large; for example, they are about 
$+40$\% at $m_T\sim 30$~GeV. This is not unexpected, 
since in this region of transverse masses, the ${\cal O}(\as)$ result corresponds
to the calculation at 
the first perturbative order and, therefore, our ${\cal O}(\as^2)$ result 
is actually only a calculation at the NLO level of perturbative accuracy.

We have illustrated 
a calculation of the cross section for $W$ and $Z$ boson production
up to NNLO in QCD perturbation theory. An analogous computation was presented
in Ref.~\cite{DYdiff}.
Our calculation is performed with a completely independent method. 
In the quantitative studies that we have carried out, the two computations
give results in numerical agreement. 
Our calculation is
directly implemented in a parton level event generator.
This feature makes it particularly suitable for practical applications
to the computation of distributions in the form of bin histograms.
A public version of our program will be available in the near future.

\noindent {\bf Acknowledgements.}
We thank Stefan Dittmaier for useful correspondence.

\end{document}